\documentclass[fleqn,usenatbib]{mnras}

\usepackage{newtxtext,newtxmath}

\usepackage[T1]{fontenc}
\usepackage{xcolor}

\DeclareRobustCommand{\VAN}[3]{#2}
\let\VANthebibliography\thebibliography
\def\thebibliography{\DeclareRobustCommand{\VAN}[3]{##3}\VANthebibliography}

\usepackage{graphicx}	
\usepackage{amsmath}	

\title[3D turbulent magneto-thermal evolution]{3D evolution of neutron star magnetic-fields from a realistic core-collapse turbulent topology}

\author[C. Dehman et al.]{
Clara Dehman,$^{1,2}$\thanks{E-mail: c.dehman@csic.es}
Daniele Vigan\`o,$^{1,2,3}$
Stefano Ascenzi$^{1,2}$, Jose A. Pons$^{4}$, Nanda Rea$^{1,2}$
\\
$^{1}$Institute of Space Sciences (ICE-CSIC), Campus UAB, Carrer de Can Magrans s/n, 08193, Barcelona, Spain\\
$^{2}$Institut d'Estudis Espacials de Catalunya (IEEC), Carrer Gran Capità 2–4, 08034 Barcelona, Spain\\
$^{3}$ Institute of Applied Computing \& Community Code (IAC3), University of the Balearic Islands, Palma, 07122, Spain \\
$^{4}$Departament de Física Aplicada, Universitat d'Alacant, Ap. Correus 99, E-03080 Alacant, Spain\\
}

\date{Accepted XXX. Received YYY; in original form ZZZ}

\pubyear{2023}

\begin{document}
\label{firstpage}
\pagerange{\pageref{firstpage}--\pageref{lastpage}}
\maketitle

\begin{abstract}
We present the first 3D fully coupled magneto-thermal simulations of neutron stars (including the most realistic background structure and microphysical ingredients so far) applied to a very complex initial magnetic field topology in the crust, similar to what recently obtained by proto-neutron star dynamo simulations. In such configurations, most of the energy is stored in the toroidal field, while the dipolar component is a few percent of the mean magnetic field. This initial feature is maintained during the long-term evolution ($\sim10^6$\,yr), since the Hall term favours a direct cascade (compensating for Ohmic dissipation) rather than a strong inverse cascade, for such an initial field topology. The surface dipolar component, responsible for the dominant electromagnetic spin-down torque, does not show any increase in time, when starting from this complex initial topology.
This is at contrast with the timing properties of young pulsars and magnetars which point to higher values of the surface dipolar fields. A possibility is that the deep-seated magnetic field (currents in the core) is able to self-organize in large scales (during the collapse or in the early life of a neutron star). 
Alternatively, the dipolar field might be lower than is usually thought, with magnetosphere substantially contributing to the observed high spin-down, via e.g., strong winds or strong coronal magnetic loops, which can also provide a natural explanation to the tiny surface hotspots inferred from X-ray data.

\end{abstract}



\section{Introduction}
\label{sec: intro}

In the last years, increasing efforts have been dedicated to magneto-hydrodynamical (MHD) simulations of core-collapse supernovae (CCSN) (e.g., \citealt{obergaulinger2014,mosta14,mosta15,bugli20,Aloy_2021,powell22,obergaulinger22}). Besides the intrinsic importance in understanding the underlying explosion mechanisms and the fundamental physics of hot dense matter \citep{janka12,obergaulinger20}, CCSN simulations are important to define the characteristics of its compact remnant, a hot proto-neutron star (PNS, e.g. \citealt{pons1999,barrere22}). The highly dynamical process consists in three main stages, as follows. (i) For about one second after the core bounce, the system consists of a relatively cool central region surrounded by a hot mantle, collapsing and radiating off neutrinos quickly, while still also accreting material. (ii) Over the next $\sim 20$ seconds, a slowly developing state of the PNS can be identified; the system first deleptonizes and heats up the interior parts of the forming star, later it cools down further through neutrino diffusion. The PNS is born extremely hot and liquid ($T \approx 10^{10}$~K), with a relatively large radius of $\sim 100$~km. (iii) After several minutes, it becomes transparent to neutrinos and shrinks to its final radius $R\sim10-14$~km \cite{burrows1986,keil1995,pons1999}. Neutrino transparency marks the birth of a NS, which starts its long-term cooling, dominated first by neutrinos and later ($t\gtrsim 10^5$ yr) by photon emission from the surface.

The NS magnetic field configuration at birth is largely unknown, and explaining how to generate strong dipolar fields able to explain the timing properties of magnetars is still an open question (see e.g. \citealt{igoshev21rev} for a review). Different scenarios were discussed in the literature to explain the origin of the magnetic field in magnetars. Magnetic flux conservation can lead to the strongest magnetic fields in the case of highly magnetized progenitors that could be formed in stellar mergers \citep{ferrario06,schneider2019,makarenko21}, although it may not explain the formation of millisecond magnetars since highly magnetized progenitors are slow rotators \citep{shultz2018}. To have both fast rotation and a strong magnetic field a possible scenario is the magnetic field amplification by a turbulent dynamo in the PNS \citep{raynaud20}. Different approaches of magneto-hydrodynamic (MHD) local or global simulations have been recently put forward to quantify the magnetic field amplification.

On one side, box simulations with simplified background fields showed the development of the magneto-rotational instability (MRI) \citep{balbus1991,akiyama2003,obergaulinger2014,rembiasz2017,Aloy_2021}.
On the other side, global simulations have explored the PNS dynamo mechanisms \citep{raynaud20,reboul2021,masada22}, including the effects of differential and rigid rotation and convection. As usual in dynamo simulations, the system reaches an equilibrium configuration in which, although the fields are not static at all, the distribution of energy over the scales (or multipoles) is statistically at (quasi)-equilibrium due to the balance between the overall forces in the magnetized fluid. In particular, the magnetic energy of the PNS spreads over a wide range of spatial scales, with non-axisymmetric and toroidal components dominating over the poloidal large-scale dipole. The total energy is compatible with magnetar-like magnetic fields, but it is dominated by the small scales. 

In this sense, the approach and results of these studies represent a major advance that improves the highly simplified pictures of static idealized equilibria known as twisted torus, often used as initial field configurations. As a matter of fact, such purely large-scale and axisymmetric configurations have a completely different spatial distribution of the magnetic energy, compared to the dynamo configurations, but they have been often used as starting point for the long-term evolution in NSs.

Here we present the first 3D coupled magneto-thermal simulation having at the same time a realistic background structure and microphysics, and an initial magnetic field topology similar (in terms of spectral distribution) to \cite{reboul2021}. The simulation is performed using \textit{MATINS} (\citealt{dehman2023} \& Ascenzi et al. in prep.) 
The latter works used the most recent temperature-dependent microphysical calculations, a star structure coming from a realistic equation of state (EOS) and the inclusion of the corresponding relativistic factors in the evolution equations. Such simulations further complements the recent efforts to describe crust-confined 3D magnetic field evolution without or with thermal evolution that used a simplified microphysical prescription described in the \textit{PARODY}-based published works \citep{wood15,gourgouliatos16,gourgouliatos20,degrandis20,degrandis21,igoshev21a,igoshev21}. In particular, our simulations confirm the general trend seen from the initially purely small-scale field of \cite{gourgouliatos20}, but using a more realistic model and performing a deeper analysis.
 
This letter is structured as follows. In Sect.~\ref{sec: 3DMT}, we briefly review the theoretical framework adopted to simulate the magneto-thermal evolution of NSs. The results of the first 3D coupled magneto-thermal evolution using \textit{MATINS} are displayed in Sect.~\ref{sec: results}. Finally, we discuss our results and we draw the main conclusions in Sect.~\ref{sec: discussion}.

\section{3D magneto-thermal evolution}
\label{sec: 3DMT}

\subsection{Basic equations}

After few minutes from birth, NSs settle to a stratified, hydrodynamical static configuration, where no convection or differential rotation operate. If they are isolated, they have then no further relevant energy source, which implies that the kinetic (i.e., rotation), magnetic and thermal energy will decay in the long term. In fact, all the rich multi-wavelength phenomenology during the active life of NSs (beamed non-thermal emission, thermal X-ray emission, transient bursts and flux enhancements), ultimately relies only on the electro-magnetic torque and re-arrangement (via long-term dissipation and sporadic abrupt re-configurations) of the PNS huge magnetic energetic inheritance.

In order to quantify the long-term evolution of NS's magnetic fields $\vec{B}$ (in the crust) and the internal temperature $T$, the MHD equations can be then reduced to only two coupled equations (see the review by \cite{pons2019} for more details), the Hall induction equation and the heat diffusion equation for the crust, which read respectively: 
\begin{equation}
    \frac{\partial \vec{B}}{\partial t} = -\vec{\nabla}\times \Big[ \eta \vec{\nabla}\times (e^{\nu}\vec{B}) + \frac{c}{4 \pi e n_e}[\vec{\nabla} \times (e^{\nu}\vec{B})] \times \vec{B} \Big]~, 
   \label{eq: induction equation} 
\end{equation}
\begin{eqnarray}
c_{V}(T)  \frac{\partial  \big( T e^{\nu} \big) }{\partial t}  &=&  \vec{\nabla}\cdot (e^{\nu} \hat{\kappa}(T,\vec{B}) \cdot \vec{\nabla}(e^{\nu}T)) + 
 \nonumber\\ 
&&  + e^{2\nu}(Q_{J}(\vec{B},T) - Q_{\nu}(\vec{B},T)) ~,
  \label{eq:thermal_evolution}
\end{eqnarray}
where: $c$ is the speed of light, $e$ the elementary electric charge, $n_e$ the electron number density, $\eta(T)=c^2/(4 \pi \sigma_e(T))$ the magnetic diffusivity (inversely proportional to the electrical conductivity $\sigma_e$), $c_{V}$ the heat capacity per unit volume, $\hat{k}$ the anisotropic thermal conductivity tensor, $Q_J$ 
and $Q_{\nu}$ the Joule heating rate and neutrino emissivity per unit volume, and $e^\nu$ is the relativistic redshift correction.
 
The system of equation must be supplemented by an equation of state, which allows to set: (i) the fixed, spherical background structure of the star (e.g., density and composition as a function of radius) which is obtained through solving the Tolmer-Oppenheimer-Volkoff equation for a given central pressure (i.e., a given total mass), (ii) the local microphysics $\eta$, $c_V$, $\hat\kappa$, $Q_\nu$. Superfluid and superconductive models for neutrons and protons respectively must also be taken into account, since they have a huge impact on the cooling timescales, via $c_V$ and $Q_\nu$ In this study, we assume the superfluid models of \cite{ho2015}. Equation of state and superfluid models have an important impact on the cooling, but they play a lesser role in the magnetic field evolution (compared for instance to the chosen initial topology).

The main couplings between $T$ and $\vec{B}$ in their evolution equations are explicitly marked as dependencies in the equations above. On one side, a main effect is the magnetic-to-thermal energy conversion via Joule heating $Q_J$. This in turn depend on $\eta(T)$, which decreases as the NS cools down (and becomes basically temperature-independent for sufficiently low temperatures when impurities dominate the scattering processes, e.g. \citealt{aguilera2008}). On the other hand, the magnetic field makes the thermal conductivity $\hat\kappa$ anisotropic, hampering the transmission of heat across the magnetic field lines and allowing important dis-homogeneities in the surface temperature distribution $T_s$, for which observations can give constraints. At a much less extent, $\vec{B}$ has also an effect on $c_V$ and $Q_\nu$.

The core and the crust (accounting together for more than $99\%$ of mass and volume) are fully considered in the thermal evolution. For the outermost envelope, where the timescales are much shorter than in the interior, we rely on the modeling of an effective function $T_s(T_b,\vec{B})$, which depends on the temperature and the magnetic field at the bottom of the envelope $T_b$ \citep{potekhin2015}. We assume for simplicity a blackbody emission from the surface, which might be a simplification \citep{potekhin15b}, but doesn't affect our conclusions.

\begin{figure*}
    \centering
    \includegraphics[width=0.45\textwidth]{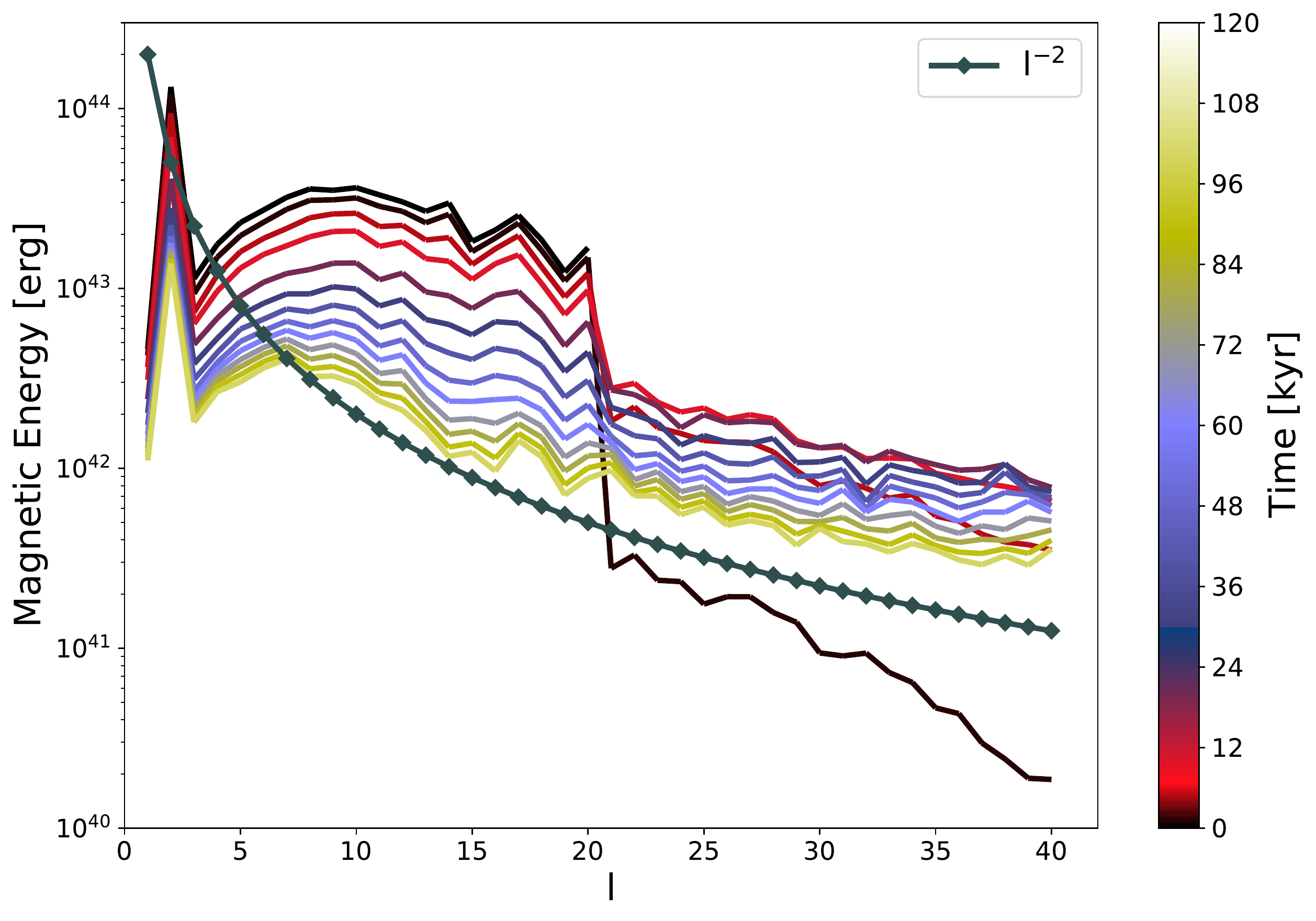}
        \includegraphics[width=0.45\textwidth]{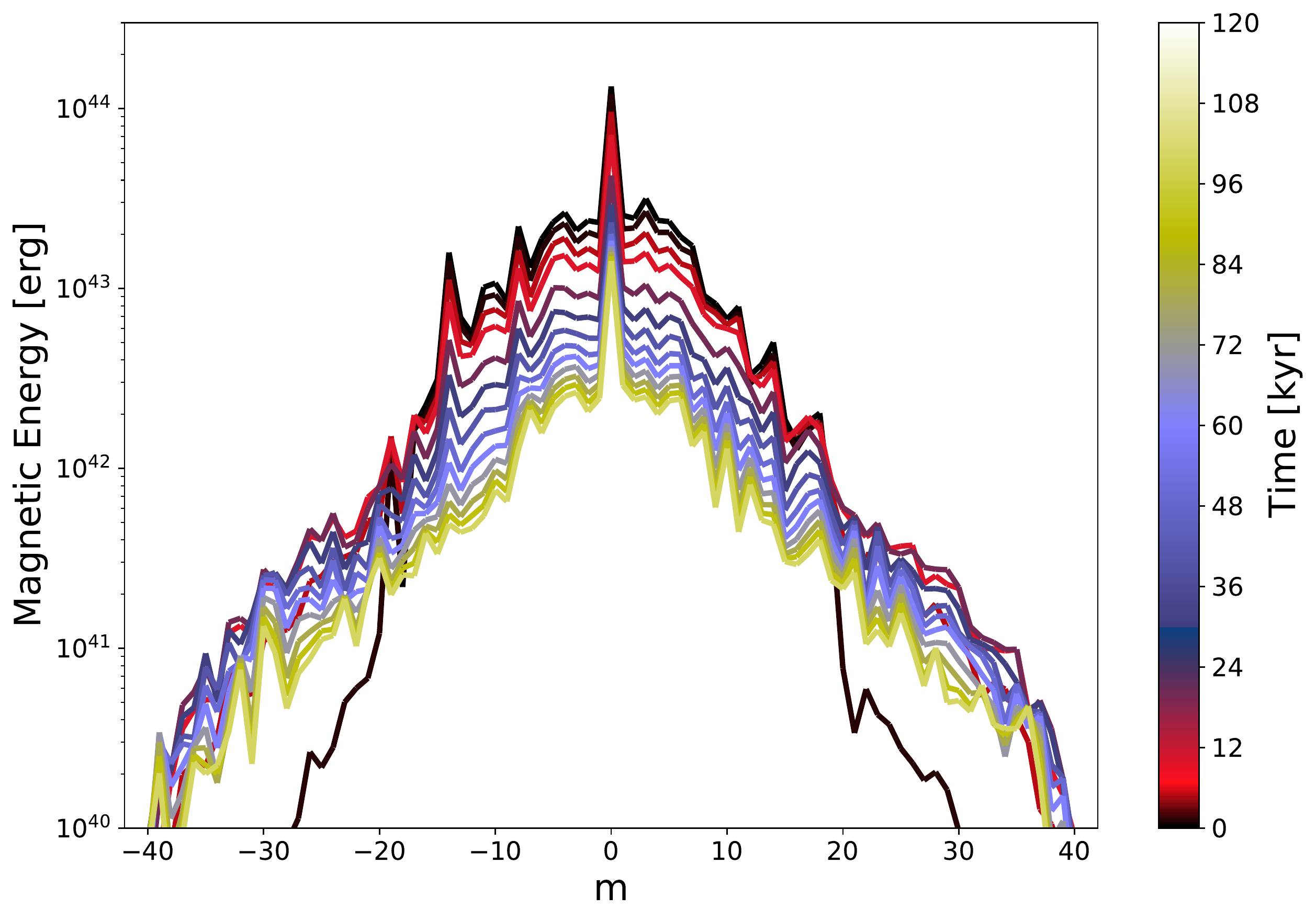}
    \caption{Spectrum of the total magnetic energy at different evolution time up to $100$~kyr (Hall balance is reached in the system). \textit{Left panel:} $l-$energy spectrum. \textit{Right panel:} $m-$energy spectrum. }
    \label{fig: energy spectrum}
\end{figure*}

We neglect the poorly known magnetic evolution in the core (see e.g. \cite{gusakov20,wood22} for recent discussions), by assuming that currents circulate only in the crust. Indeed, eq.~(\ref{eq: induction equation}) is valid for the solid crust only: the first term is the Ohmic term, and the second is the non-linear Hall term, which is the effect of the Lorentz force acting on the ultra-relativistic electrons which carry the charge by freely moving through the solid (or plastic) ion lattice.
The Hall term tends to push the electric currents toward the crust-core boundary, where the high impurity content and pasta phases could cause a fast dissipation of the magnetic field and therefore much less spin-down \cite{pons2013}. In addition, the Hall term tends to redistribute the magnetic energy across all scales. This effect compensates the Ohmic dissipation of the smallest scales and, after ${\cal O}(10^4)$ yr, the crustal magnetic topology approaches a very slowly varying configuration, with the energy distributed over a broad range of multipoles, a Hall cascade with slope $\sim l^{-2}$ for the intermediate and small scales \citep{dehman2023}.

\begin{figure}
    \centering
        \includegraphics[width=0.45\textwidth]{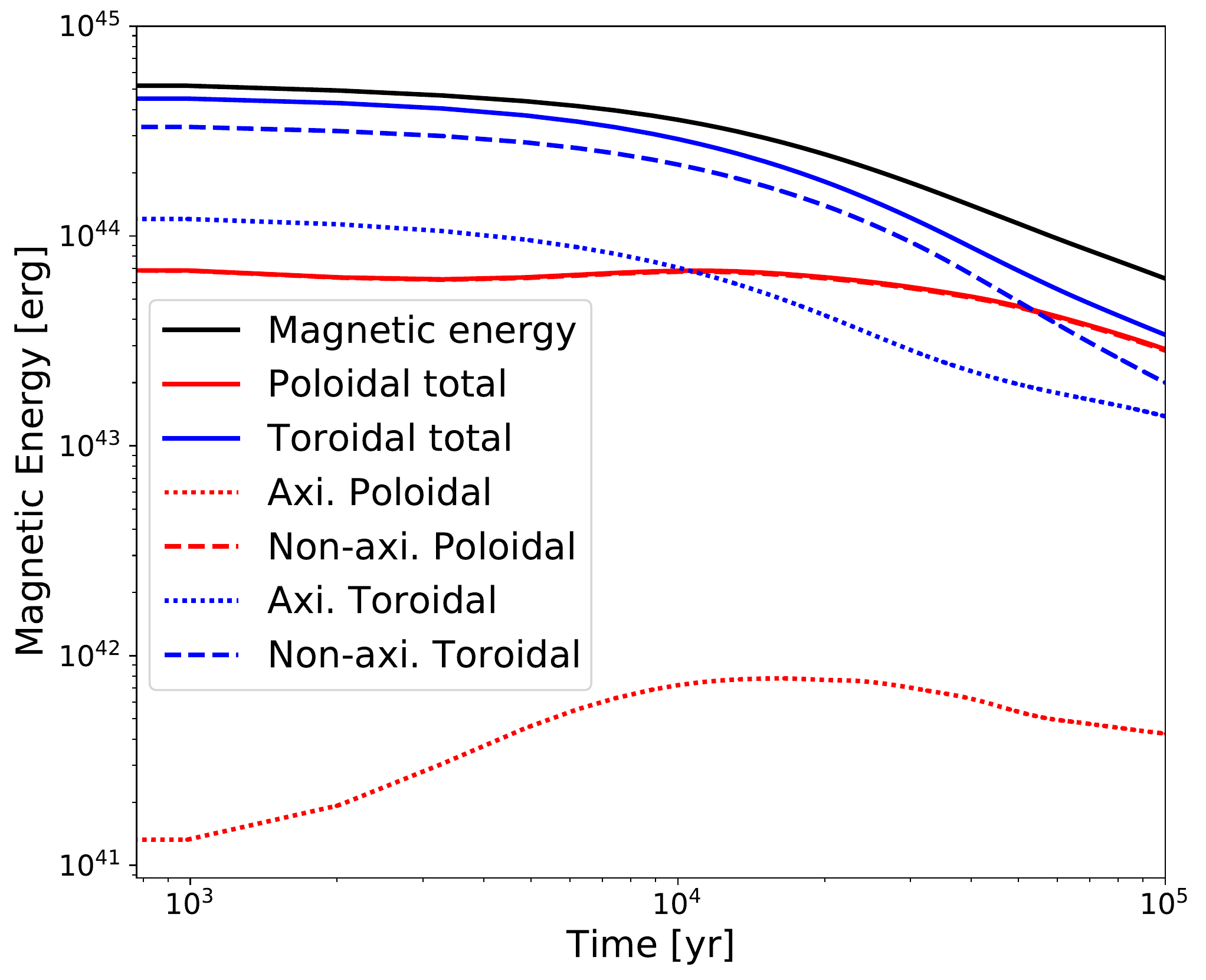}
    \caption{Decomposition of the poloidal and toroidal magnetic energy as a function of time. Poloidal magnetic energy is represented with red color, the toroidal energy with blue, and the total magnetic energy with black solid lines. The dots correspond to the axisymmetric components of these energies and the dashed lines to the non-axisymmetric components. The axisymmetric and non-axisymmetric component are taken in PNS dynamo simulations with respect to the axis of the PNS rotation. This has no role in our simulations and we define these components with respect to the magnetic axis, starting with only $m=0$ for a dipole.}
    \label{fig: Emag PT}
\end{figure}

\begin{figure*}
    \centering
    \includegraphics[width=0.47\textwidth]{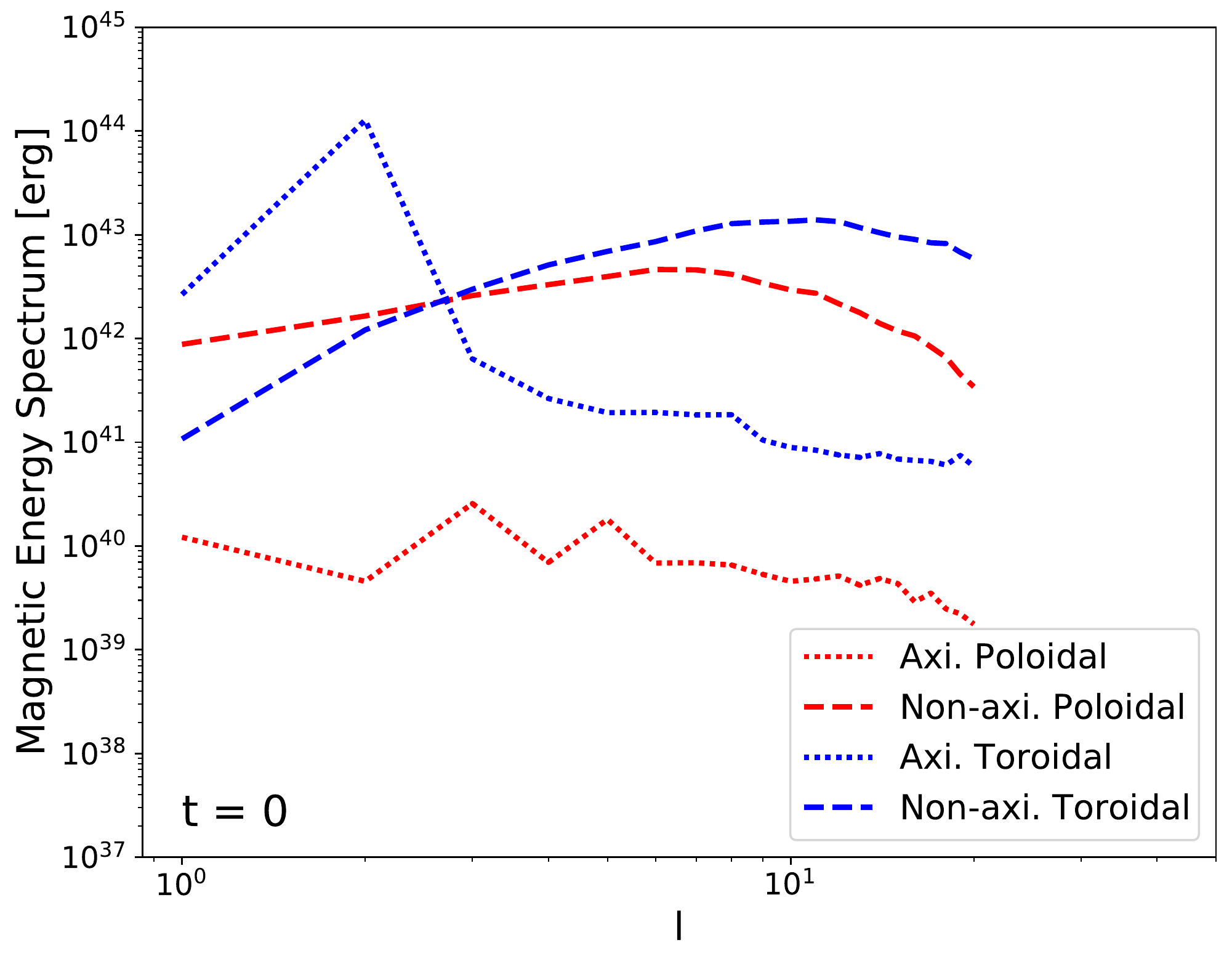}
    \includegraphics[width=0.47\textwidth]{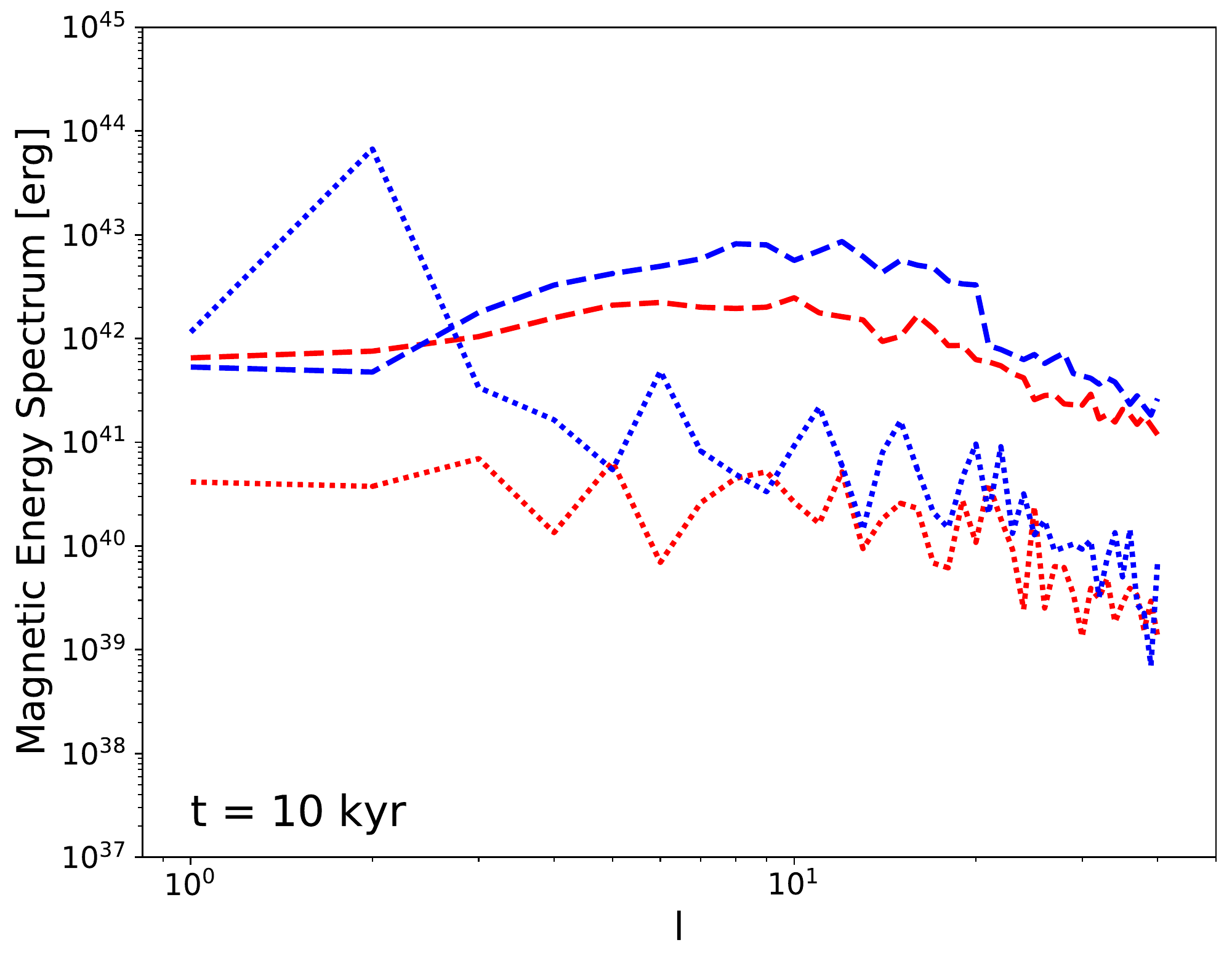}
    \includegraphics[width=0.47\textwidth]{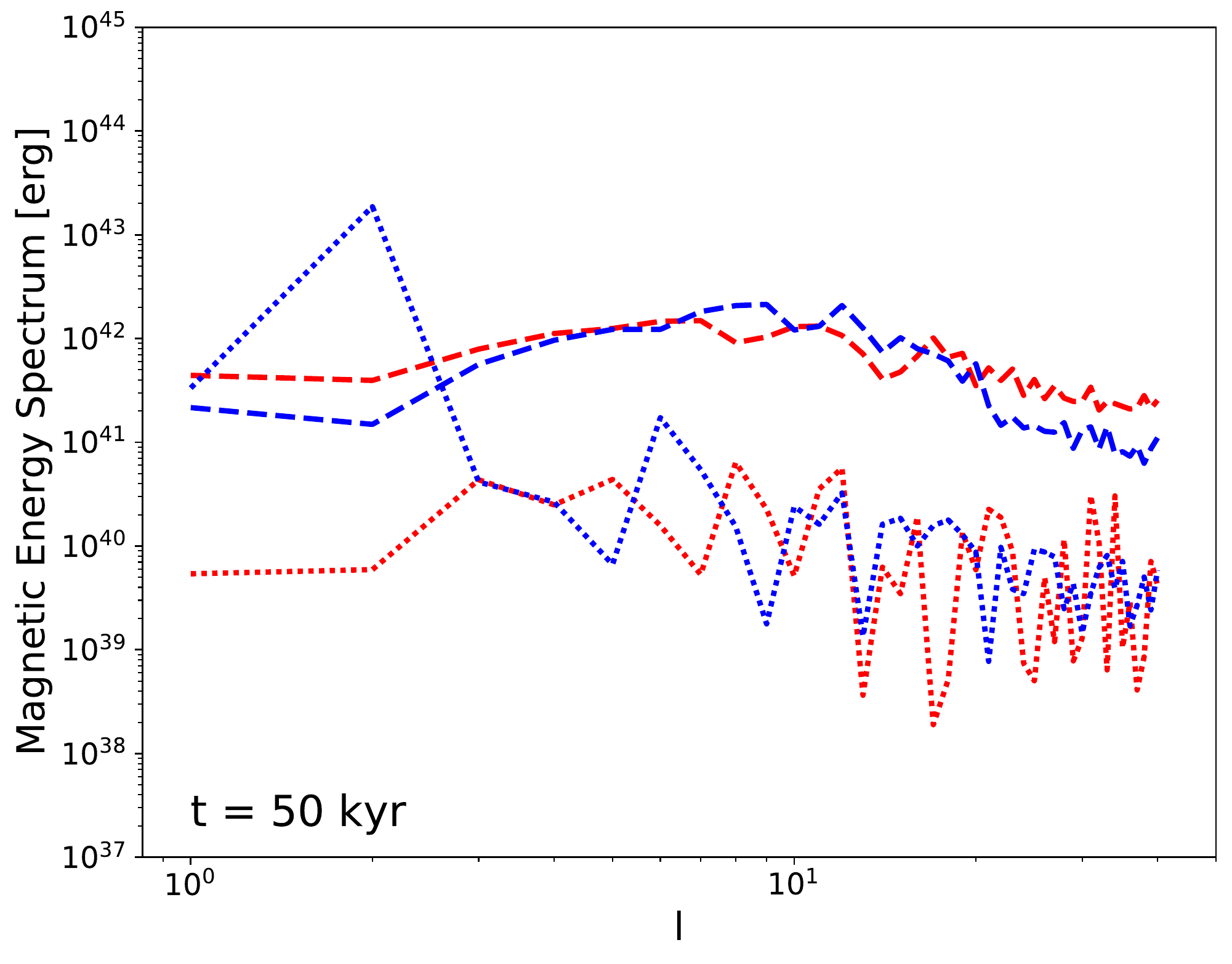}
    \includegraphics[width=0.47\textwidth]{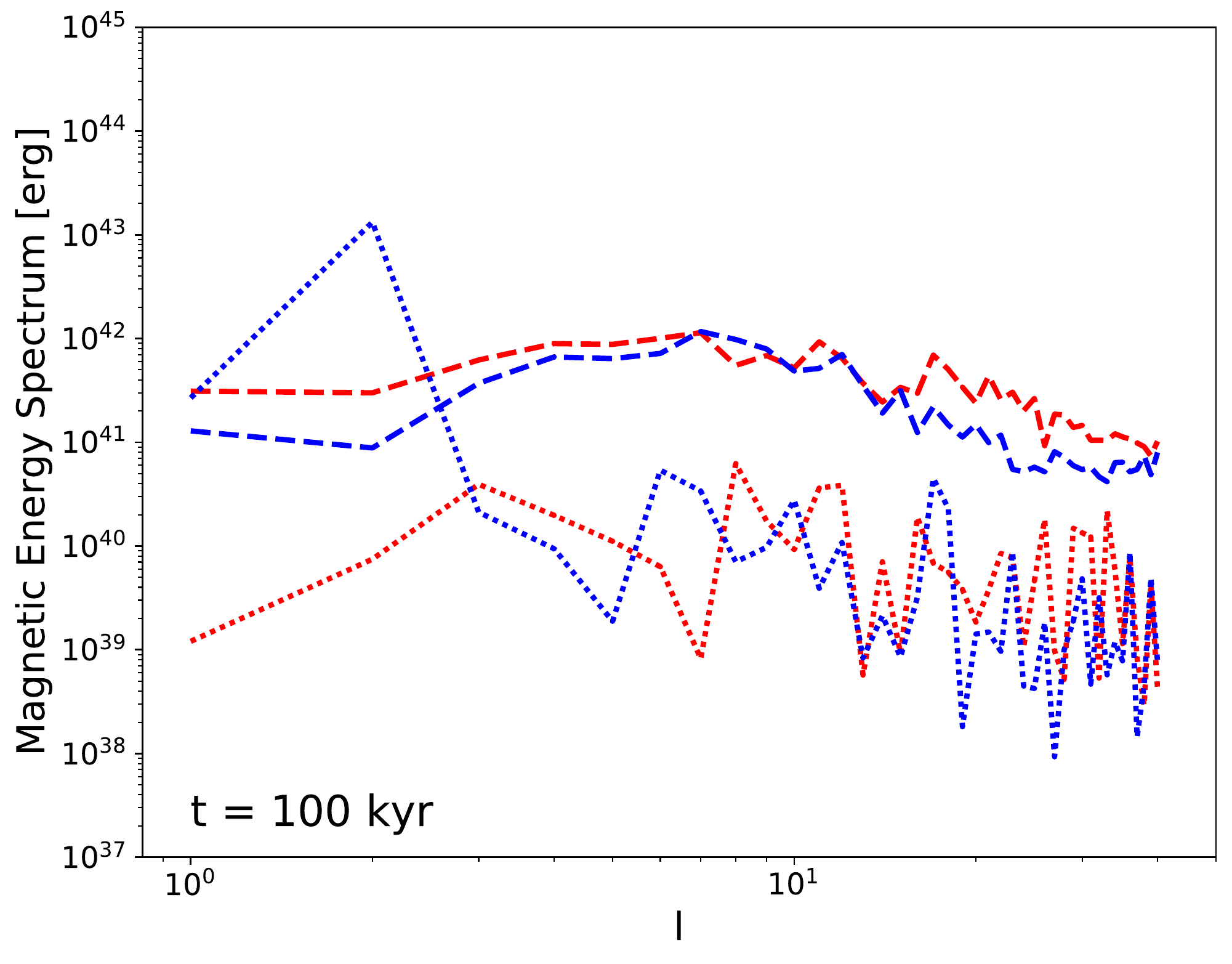}
    \caption{ Spectrum of the toroidal magnetic energy (blue) and the poloidal magnetic energy (red) as a function of the spherical harmonics order $l$ at $t=0, 10, 50$ \& $100$~kyr. The dots correspond to the axisymmetric components of these energies and the dashed lines to the non-axisymmetric components. The total number of multipoles in the system is $l_{\tt max}=40$ and the initial number of multipoles is $l=20$. The top left panel of this figure ($t=0$) is inspired from a core-collapse generated turbulent field \citep[Fig.~7 top panel]{reboul2021}. }
    \label{fig: TP}
\end{figure*}

\subsection{Numerical setup}
We use \textit{MATINS} \citep{dehman2023}, a finite-volume 3D code employing the cubed-sphere coordinates, originally introduced by \cite{ronchi96}.
\textit{MATINS} is designed for this scenario and evolves both the Hall induction and the heat diffusion equations 
eqs.~(\ref{eq: induction equation}) and (\ref{eq:thermal_evolution}) 
(see \citealt{dehman2023} and Ascenzi et al. in prep. for details about the magnetic and thermal evolution details separately).
We consider the magnetic field confined in the crust employ potential magnetic boundary conditions (current-free magnetosphere) at the outer numerical boundary, places at density $\rho=10^{10}$ g~cm$^{-3}$ (close to the transition between liquid envelope and solid crust takes place for young or middle-age NSs). We implement the state-of-the-art calculations for the temperature-dependent electrical conductivity at each point of the star using Potekhin's public codes\footnote{\url{http://www.ioffe.ru/astro/conduct/}} \citep{potekhin2015}. The background NS model can be built using different models of the equation of state at zero temperature, taken from the online public database CompOSE\footnote{\url{https://compose.obspm.fr/}} (CompStar Online Supernovae Equations of State). In particular, we will show results that employ a Skyrme-type model of EOS, SLy4 \citep{douchin2001}, and we consider a mass $M=1.4 M_{\tt sun}$. We employ the $T_s(T_b,\vec{B})$ relation of \cite{potekhin2015} for iron magnetized envelopes. For more details on the impact of magnetized envelopes on the cooling of NSs, we refer the reader to \cite{dehman2023b}. The microphysics, star's structure and envelope modules are the same as in our 2D models \citep{vigano2021}.

We adopt a grid resolution of $N_r=40$ and $N_\xi=N_\eta=43$ per patch (a cubed-sphere has $6$ patches), corresponding to a total number of resolved multipoles in the system of about $l_{\tt max} \sim 40$. We follow the evolution for 100 kyr, an age where most NSs have cooled down enough to be hardly detectable (bolometric thermal luminosity $L < 10^{32}$~erg~s$^{-1}$).


\subsection{Initial topology}
\label{sec: initial topology}

We start from the work by \cite{reboul2021}, who performed global simulations using MagIC \citep{wicht2002}. They consider a shell representative of the convective region of a PNS, and simulate the dynamo coming from the typical differential rotation profile seen during an advanced state of the core-collapse. At saturation, they found a very complex topology, with most of the magnetic energy contained in the toroidal axisymmetric large-scale component (especially the quadrupolar component given by winding) and in the non-axisymmetric small or medium-scale size magnetic structures, both for the toroidal and the poloidal components. The large-scale dipolar component represents only about $\sim 5\%$ of the average magnetic field strength.

We adopt an initial field with an angular spectral energy distribution, shown in the upper left panel of Fig.~\ref{fig: TP}, comparable to \citep[Fig.~7 top panel]{reboul2021}. Here we indicate toroidal (blue)/poloidal (red) and axisymmetric (spherical harmonics order $m=0$)/non-axisymmetric ($m\neq 0$) components as a function of the spherical harmonics degree $l$. The initial configuration is then a smooth cascade in $l$ (truncated for simplicity at $l=20$) and $m$, with an exception for the quadrupolar toroidal component which dominates. The average initial magnetic field of a few $10^{14}$ G, corresponds to a total magnetic energy of the order of $\sim 6 \times 10^{44} $ erg. Note that axial symmetry here refers to the rotational axis in the PNS phase, but the rigid rotation of the NS plays no role in the magnetic evolution (we don't evolve the full MHD, with the momentum equation, that includes the Coriolis force, since the background is assumed static for the solid crust). Therefore, hereafter, in our simulations, the reference axis can be thought as the PNS rotational axis.

In the absence of a realistic core evolution, we simplify the topology by confining the magnetic field to the NS crust (smaller than the PNS shell), using the same (arbitrary) initial radial dependence described in \cite{dehman2023}. If the core evolution timescales are much longer than the crust (see e.g. \citealt{gusakov20}), most results of the crustal evolution should hold anyway.

Of course, it is unrealistic to expect that the birth NS configuration is given exactly by the ones simulated for the PNS stage. As a matter of fact, there is plenty of MHD timescales to change the topology. On one side, PNS will experience shrinking to the final NS size, which will probably imply an increase of the energy by flux conservation (here considered since we obtain similar volume-integrated energy spectra). On the other side, the smaller scale will experience a fast decay as soon as the dynamo processes will stop feeding them. This is the reason why, to be conservative, we only consider the first $l\leq 20$ multipoles. However, the details of the spectral slope for the small scales are not important, since the Hall effect quickly regenerates them via direct cascade \citep{dehman2023}.

\section{Results}
\label{sec: results}

In this section, we present the results of our simulation assuming a crustal-confined initial field topology as described in Sect.~\ref{sec: initial topology}. The studied model has an average magnetic field of a few $10^{14}$ G and a dominant Hall evolution (the magnetic Reynolds number is greater than unity and reaches a maximum of $R_m \sim 1600$ during evolution). For a more quantitative analysis of the 3D magnetic evolution, we survey the magnetic energy spectrum to observe the redistribution of the magnetic energy over the different spatial scales.

\begin{figure*}
\centering
\includegraphics[width=0.9\textwidth]{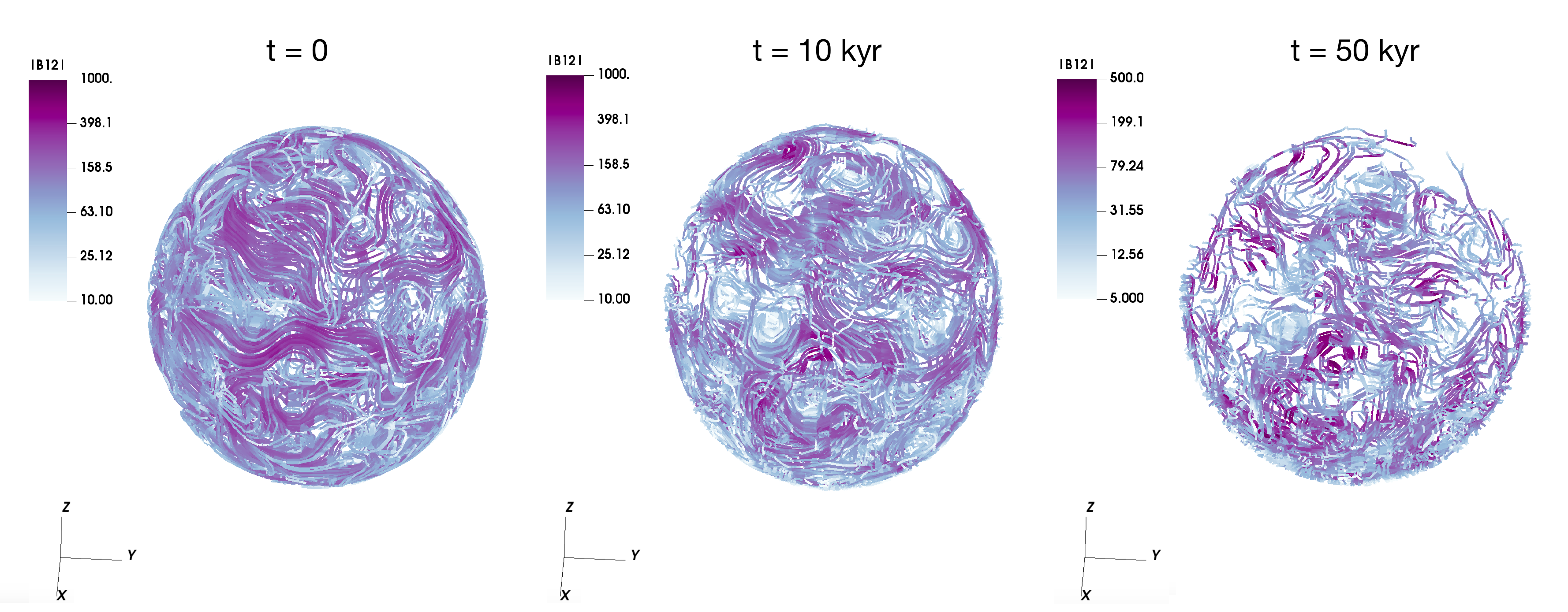}
\caption{Field lines in the crust of a NS at $t=0$ (on the left), $t=10$~kyr (in the center) and $t=50$~kyr (on the right). The color scales indicates the local field intensity in units of $10^{12}$~G.}
\label{fig: field lines visit}
\end{figure*}

We examine at different evolution times the $l$ and $m$ energy spectrum, illustrated respectively in the left and right panels of Fig.~\ref{fig: energy spectrum}. The transfer of the magnetic energy occur from large to small scales (the latter are initially empty): the direct cascade dominates. The magnetic energy spectra have reached a quasi-stationary state, i.e., the Hall-saturation, at about $20-30$~kyr. 
Small-scale structures dissipate faster than large-scale ones, enhancing the Ohmic heating in the system. At the same time, the former are continuously fed by the latter, thanks to the Hall term in the induction equation. This is known as the Hall cascade, it happens thanks to the Hall-dominated dynamics and it consists in an equilibrium distribution of magnetic energy, over a quite broad range of multipoles, with an approximate $l^{-2}$ slope \citep{goldreich1992}. 

Throughout $\sim100$~kyr of evolution, the total magnetic energy drops by about half an order of magnitude as indicated in Fig.~\ref{fig: Emag PT} (solid black line). We also display in the same figure the decomposition of the total magnetic energy (black) into its poloidal (red) and toroidal (blue) components. The dots correspond to the axisymmetric components ($m=0$) and the dashed lines to the non-axisymmetric ones ($m\neq0$). The studied initial topology has a poloidal field governed by its non-axisymmetric component whereas the toroidal field has a significant contribution from both axisymmetric and non-axisymmetric pieces. For the first $\sim 10^4$~kyr, the dissipation of magnetic energy is relatively small compared to later time. One could notice that the magnetic energy is transferred from the toroidal to the poloidal field during the evolution. The most efficient transfer in terms of relative energy increase, is for the poloidal axisymmetric field, probably because it is initially much weaker than the others. It shows a significant growth of one order of magnitude during the first $40-50$~kyr. A saturation occurs for rest of the evolution (i.e., the system has reached the Hall balance). This quasi-constant energy trend appears for both the poloidal and the toroidal axisymmetric energy for $t \leq 50$~kyr. At the same time, a strong transfer of magnetic energy to the poloidal non-axisymmetric component takes place. The non-axisymmetric toroidal energy tends to dissipate faster than the total magnetic energy in the system. That is because part of the toroidal non-axisymmetric is transferred to the poloidal non-axisymmetric energy. The latter dominates the axisymmetric toroidal component at $\sim 10$~kyr and the non-axisymmetric one at $\sim 50$~kyr. Nevertheless, the total toroidal field governs the magnetic energy at all time (solid blue line).

For a further understanding, we study in Fig.~\ref{fig: TP} the evolution in time of the spectra. We display four snapshots of the toroidal (blue)/ poloidal (red) and axisymmetric (dots)/non-axisymmetric (dashed lines) components as a function of the spherical harmonics degree $l$ at different evolution time, e.g., $t=0, 10,50$ \& $100$~kyr. 
At $\sim10$~kyr, a transfer of energy to the non-axisymmetric toroidal dipole (blue dashed line, $l=1$) takes place due to the inverse Hall cascade. A significant transfer of energy to the poloidal axisymmetric field happens at large- and small-scales. Instead, at late time (Hall balance is reached in the system) 
the dipolar component dissipates. For a better understanding, we discuss the behaviour of small- and large-scales independently. 
During the evolution, the small-scale modes ($ 10 \leq l \leq 40$) gain a significant fraction of the magnetic energy thanks to the Hall cascade in the system. The different behaviors are listed below
\begin{itemize}
    \item At $t=10$~kyr, both toroidal components (i.e., axisymmetric and non-axisymmetric) decay in time, whereas the two poloidal components gain energy at small scales. That also agrees with Fig.~\ref{fig: Emag PT} and indicates equipartition at small scales, since isotropy is easier to achieve. As a matter of fact, the peculiar crust geometry (a thin shell of $\sim 1$~km) and the strong stratification, play against isotropy and, therefore, equipartition between large-scale components.    
    \item At about $40-50$~kyr, the system approaches an Hall-balance (see also Fig.~\ref{fig: energy spectrum}). At this stage, one can notice a slightly different behaviour for the axisymmetric and the non-axisymmetric components. On one hand, the axisymmetric components reach an approximate equipartition of the magnetic energy between poloidal and toroidal components at small-scales. That is in agreement to what was found using the axisymmetric 2D code \cite{pons2019}. A change of phase in the oscillations occur on a timescale of $40-50$~kyr. On the other hand, for the non-axisymmetric modes, the slight toroidal dominance over poloidal seen at $10$~kyr inverts at $50$ and $100$~kyr. That is due to a transfer of energy from toroidal to poloidal field which, however, can be interpreted as equipartition of energy on the small isotropic scales. 
\end{itemize}
Minor differences appear in the spectrum at $50$ and $100 $~kyr and that is because the system has reached the Hall balance and the spectra remains stationary as it is shown in Fig.~\ref{fig: energy spectrum}. Note also how the large scales barely evolve, compared to the others. This is another confirmation that the NS tends to a universal behaviour (Hall cascade) for intermediate and small scales, but it has a strong memory of the large-scale magnetic topology at birth \citep{dehman2023}. This has important implications to relate current observables to the formation process (PNS stage).

The field lines in the crust of a NS at $t=0$, $10$ and $50$~kyr are displayed in Fig.~\ref{fig: field lines visit}. The magnetic field lines are highly multipolar and many small scale structures cover the surface. The field lines are very tangled throughout evolution, making it difficult to discern any clear dominant component.  

\begin{figure}
    \centering
      \includegraphics[width=0.45\textwidth]{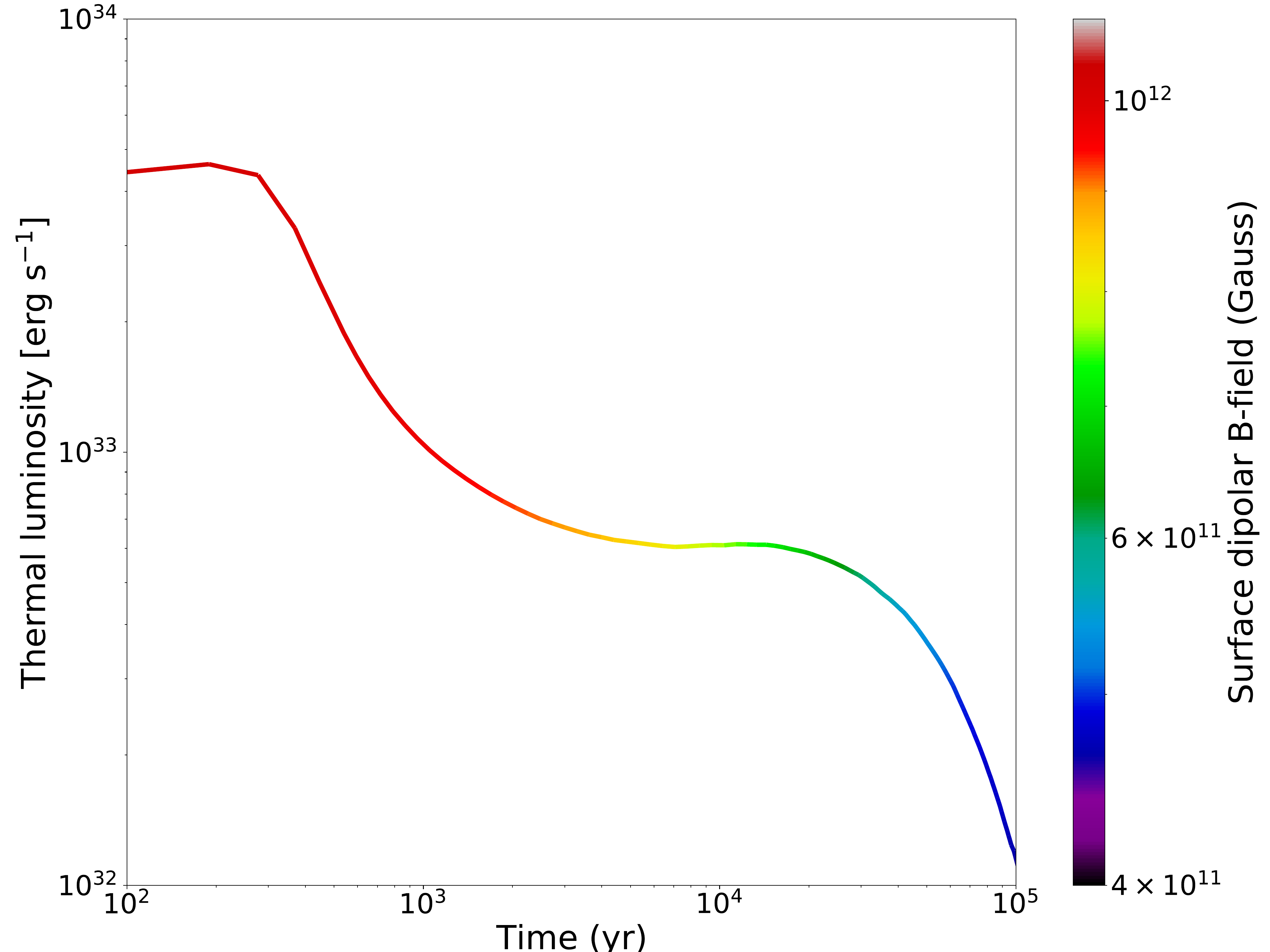}
    \caption{Luminosity curve as a function of time.   The colorbar indicates the evolution of the dipolar component of the poloidal magnetic field at the surface of the star (value at the magnetic pole). }
    \label{fig: luminosity bdip alpha}
\end{figure}

In Fig.~\ref{fig: luminosity bdip alpha}, we show the evolution of the thermal luminosity and (in color scale) of 
$B_{\tt dip}^{\tt pol}$, the dipolar component of the poloidal magnetic field at the surface of the star (value at the magnetic pole). 
For such an initial configuration, the luminosity ranges from $5\times10^{32}-10^{33}$~erg/s in the neutrino cooling era, soon after, in the photon cooling era (t$\sim 10^5$)~yr, the luminosity drops sharply below $10^{32}$~erg/s.
The rapid cooling during the photon cooling era is also caused by the low core heat capacity, which in turn depends on the assumed pairing details. A comprehensive revision of the microphysics embedded in magneto-thermal models can be found in \cite{potekhin2015}. On the other hand, $B_{\tt dip}^{\tt pol}$ drops from $\sim 10^{12}$~G to $\sim 4\times 10^{11}$~G without any noticeable increase. 

\begin{figure}
\centering
\includegraphics[width=0.45\textwidth]{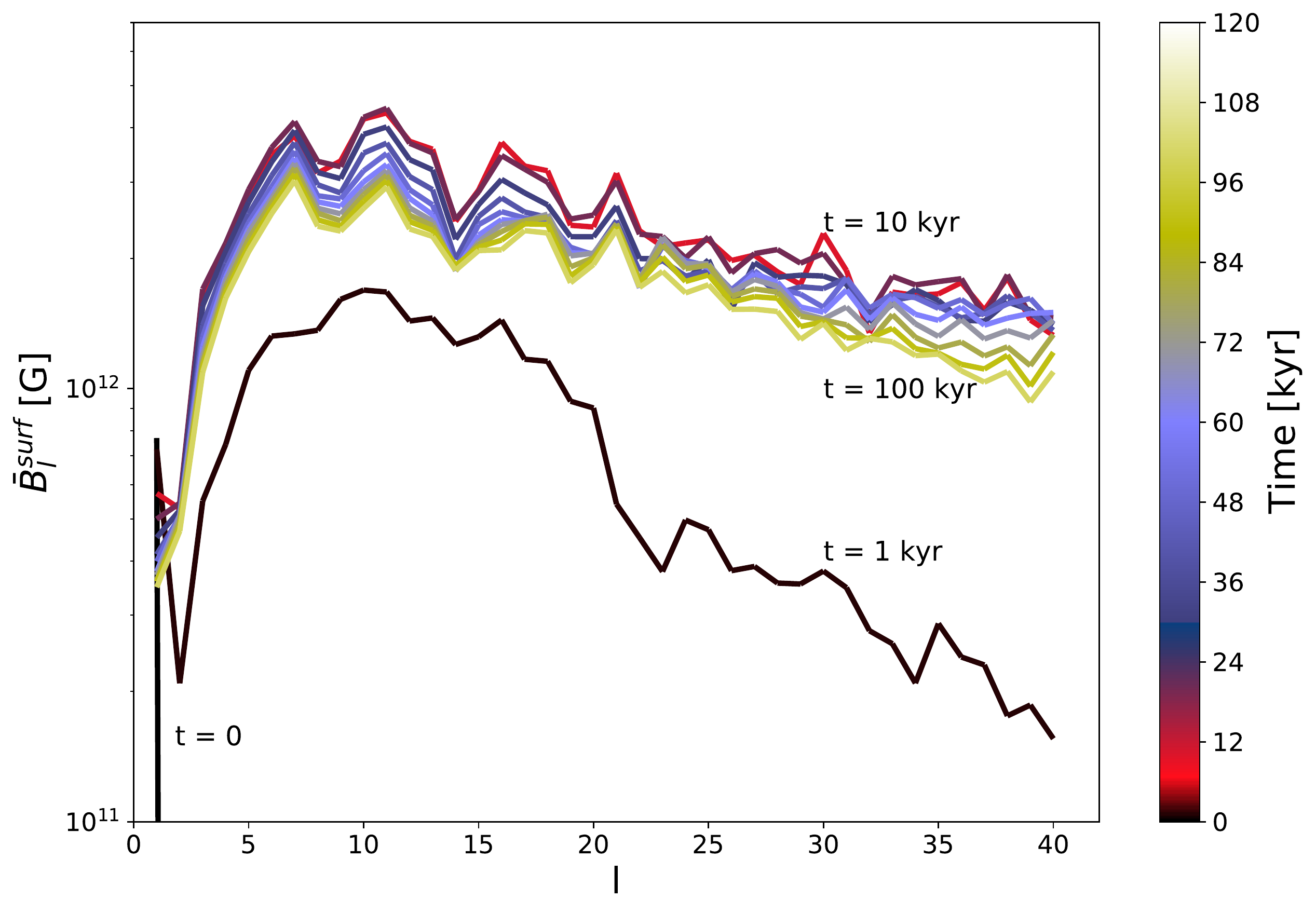}
\includegraphics[width=0.45\textwidth]{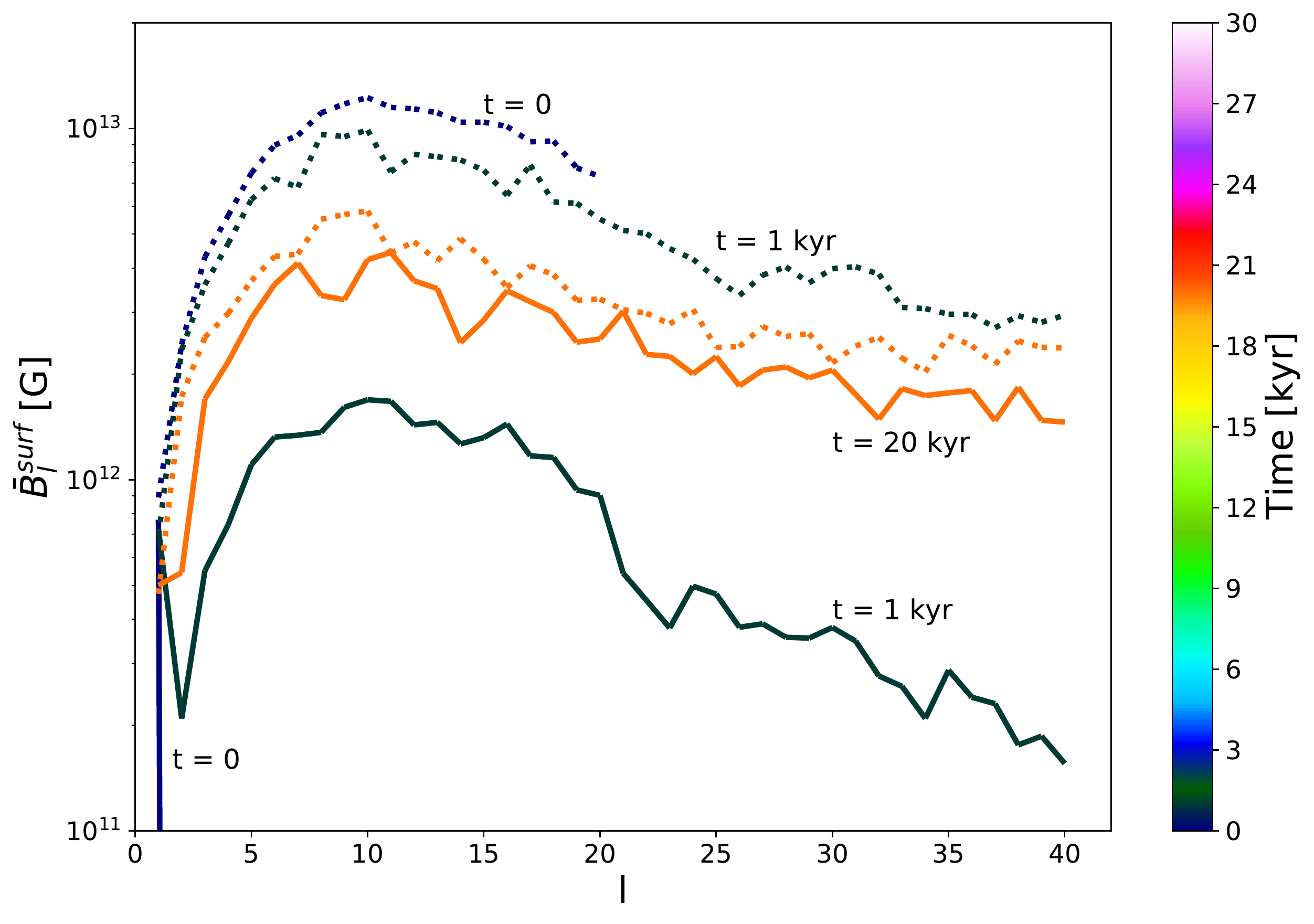}
\caption{Time evolution of the surface average field strength as a function of $l$ (eq.~\ref{eq: B surface}).
\textit{Top Panel:} evolution up to $100$~kyr. The colorbar reflects the evolution in time. \textit{Bottom Panel:} Comparison of the evolution for different radial functions at $t=0$ (black), $1$~kyr (red) and $20$~kyr (yellow). The solid lines correspond to the radial function that fits the potential boundary conditions \citep[, App.~B]{dehman2023} and results in an initial purely dipolar field at surface (black solid lines), whereas the dots correspond to \citep[eq.~8]{aguilera2008} applied to both toroidal and poloidal field.
The latter allows an initial distribution of small- and large-scale multipoles on the star's surface.}
\label{fig: blm surface}
\end{figure}

At the surface of the star, we define the average field strength in each multipole $l$ as follows 
\begin{eqnarray}
 \bar{B}^{\tt surf}_l &=&  \bigg[ \frac{1}{4\pi} \int d\Omega (B_r^2+ B_\theta^2 + B_\phi^2) \bigg]^{0.5}               \nonumber\\
 &=&\bigg[ \frac{B_0^2}{4\pi} \sum_m (b^m_l)^2  ~ \bigg( e^{-2\lambda(R)} (l+1)^2  + l(l+1)\bigg)\bigg]^{0.5},~ 
    \label{eq: B surface}
 \end{eqnarray}
where $e^{-2\lambda(R)}$ is the relativistic metric correction at the surface, $b_l^m$ are the dimensionless weights of the multipoles entering in the spherical harmonics decomposition of the radial magnetic field, and $B_0$ is the normalization used in the code (see eq.~26 of \citealt{dehman2023} for more details). 
In the top panel of Fig.~\ref{fig: blm surface}, we show the time evolution of $\bar{B}^{\tt surf}_l$ as a function of $l$.  
At t=0, the NS surface is dominated by the dipolar mode due to the specific radial function that we assume for each multipole in this case. However, as soon as we start the evolution the small-scale modes become dominant at about $1$~kyr. Then the system reaches some sort of balance throughout the evolution. The balanced configuration is dominated by small-scale structures only ($l\geq 10$). This finding implies that during the evolution the $B_{\tt dip}^{\tt pol}$ value (the colorbar of Fig.~\ref{fig: luminosity bdip alpha}) is much weaker than the one inferred for most magnetars ($10^{14}-10^{15}$ G), when the classical spin-down formula is used to interpret the timing properties.

So far we have focused only on different choices of the tangential distribution of magnetic energy (the initial multipole weights). However, the initial radial profile of the topology (i.e., the set of radial functions of each multipole $(l,m)$) is also an important parameter that potentially affect our results (also connected to the outer and inner boundary conditions). For this reason, in the bottom panel of Fig.~\ref{fig: blm surface}, we compare the evolution for two different initial sets of radial functions (both equally arbitrary), which result in very different initial shapes of $\bar{B}^{\tt surf}_l$. The solid lines correspond to a set of radial functions that allow a smooth matching with a pure dipole outside (see App.~B of \citealt{dehman2023}), confining inside all the other multipoles. The dots correspond to another set of radial functions that allows an initial distribution of small- and large-scale multipoles on the star's surface, with an initial non-zero tangential current that is quickly dissipated by the re-arrangement of the field. Both simulations are soon ($\sim 1$~kyr, red) dominated by the small-scale structures and approach a similar spectral distribution of $\bar{B}^{\tt surf}_l$ after about $20$~kyr, when the Hall balance is reached. From the comparison of these two simple choices, it seems that, independently from the initial radial distribution of the magnetic field inside the NS's crust, the star surface is anyway dominated by small-scale structures.

\section{Discussion}
\label{sec: discussion}

In the previous work \citep{dehman2023}, we have explored different initial field configurations using \textit{MATINS}. Our simulations confirmed that the spectra and topology of the magnetic field keep a strong memory of the initial large scales, which are much harder to be restructured or created. This indicates that the type of large-scale configurations attained during the PNS stage and the NS formation is crucial to determine the magnetic field topology at any age of its long-term evolution. Dynamo simulations of PNS show a complex field configuration in which most of the energy is stored in the toroidal field and only a small fraction ($\sim 5\%$) of it is stored in the dipole component, with small-scale components dominating over the large-scales ones (in particular, the dipolar poloidal component responsible for the spin-down).

To assess how such complex initial topology evolves, we presented the first coupled 3D magneto-thermal simulation of a NS field evolution, starting from a configuration similar (in energy spectra) to the recent PNS dynamo simulations by \cite{reboul2021}. We include the most realistic background structure and microphysical ingredients so far. We perform a long-term simulation until a age of $\sim 100$~kyr, i.e., until the Hall balance is reached and the luminosity evolution is driven by mostly by photon cooling.

Following the analysis of our results (Sect.~\ref{sec: results}), we found that the surface dipolar component experiences no relevant growth in time in this timescale, independently from the initial radial distribution of the magnetic field in the NS crust (see Fig.~\ref{fig: blm surface}, bottom panel). We argue that only starting from an initial magnetic energy distribution mostly concentrated in the dipolar component alone (in the crust or in the core) {\it could} result in a dominant surface dipolar magnetic field. As in our previous work, and in PARODY-based studies \citep{gourgouliatos20,igoshev21}, we find a wide range of spatial scales over which the magnetic energy is distributed, and the large-scale components are sub-dominant. However, it is unclear how core collapse could yield to a NS with such an almost pure dipolar configuration. Indeed, several observations suggest that internal non-dipolar components (toroidal and multipolar components) are dominant: low field magnetars \citep{rea2013,tan2023}, high-B pulsars \citep{zhu2011}, CCO with outbursts \citep{rea2016}, spectral features in magnetars \citep{tiengo2013} and X-ray Dim Isolated NSs (XDINS) \citep{borghese2017}. At last, in \cite{dehman2020}, the 2D simulations showed that $B_{\tt dip}^{\tt pol}$ played a minor role in determining magnetars bursting activity and the magnetic energy stored in the crust of the star is a better indicator.

These features, already partially explored in \cite{gourgouliatos20,igoshev21} (where they started with only small-scale multipoles), could be suitable for describing CCOs, X-ray sources with luminosity ranging between $10^{32}-10^{34}$~erg/s, located at the centres of supernova remnants, with an estimated surface dipole magnetic field in the range $10^{10}-10^{11}$~G. The dissipation of such weak large-scale component alone cannot provide sufficient thermal energy to power their observed X-ray luminosity. It is believed instead that CCOs have a hidden strong magnetic field due to the fall-back accretion \citep{ho2011,vigano2012}. This magnetic field dissipates in the star interior to provide the bright thermal luminosity of CCOs (Fig.~\ref{fig: luminosity bdip alpha}).



We may consider that some other physical effects could act during the collapse or in the early NS life, to provide at the surface of the star large-scale dipole $B_{\tt dip}^{\tt pol}\gg 10^{13}$ G. Strong inverse cascade could be triggered by helical magnetic fields (e.g. \citealt{brandenburg2020} for box simulations with no stratification), but it has not been seen so far in global simulations. 
We might be missing some important dynamics able to have a large-scale organization of the sustaining currents.
On one side, the PNS dynamo simulations are intrinsically affected by the chosen simple boundary conditions (see e.g. \cite{raynaud20} for a comparison between perfect conductor and potential configuration), which are anyway not realistic for the dirty, dense and hot plasma-filled environment of the PNS. On the other hand, the field sustained by currents in the core might be relevant, especially in the long-term, since its evolution timescales are arguably longer (but see \citealt{gusakov20}). However, it is unclear how to generate a strong poloidal dipolar field (supported by strong toroidal and organized currents) during the PNS or early NS stages.

A radically different alternative is possible, compatible with the absence of strong dipolar fields. The values of $B_{\tt dip}^{\tt pol}$ might indeed always be lower if the electromagnetic torque is dominated by magnetospheric effects, like particle winds and the presence of strong and extended loops, charges of particles. Indeed, for the Sun \citep{yeates2018} and for Zeeman-Doppler studies of main sequence stars \citep{neiner2009} inferred magnetospheric topology show field lines stretched by the wind and complex topologies. Strong magnetospheric structures could also be compatible with the presence of tiny hotspots (radii less than $1$~km), commonly inferred from magnetars' X-ray thermal spectra (in quiescence or outburst), and with the more studied resonant Compton scattering causing the observed non-thermal tails in spectra. However, quantifying the additional magnetospheric contributions to the spin-down are so far limited by the models of \cite{tong2013}, and more work is needed in this sense to support this scenario. Along this line, the coupling of the interior evolution with the magnetosphere is essential to allow the currents to flow and the surface dipolar field to be larger \citep{akgun2018,urban2023}.

\section*{Acknowledgements}
CD has the partial support of NORDITA while working on this article. 
CD, SA and NR are supported by the ERC Consolidator Grant 'MAGNESIA' No. 817661 (PI: N. Rea) and this work has been carried out within the framework of the doctoral program in Physics of the Universitat Aut\`onoma de Barcelona and it is partially supported by the program Unidad de Excelencia Mar\'ia de Maeztu CEX2020-001058-M. 
DV is supported by the European Research Council (ERC) under the European Union's Horizon 2020 research and innovation programme (ERC Starting Grant 'IMAGINE' No. 948582, PI: DV).
JAP acknowledges support from the Generalitat Valenciana grants ASFAE/2022/026 (with funding from NextGenerationEU PRTR-C17.I1) and the AEI grant PID2021-127495NB-I00.

\section*{Data Availability}
 
All data produced in this work will be shared on reasonable request to the corresponding author.



\bibliographystyle{mnras}
\bibliography{mnras_template} 



\bsp	
\label{lastpage}
\end{document}